\documentclass[12pt]{article}
\usepackage{graphicx}
\begin{document} 

\centerline{\large Experimental realities refuting the existence of p=0
condensate in a system} 

\centerline{\large of interacting bosons : II. Spectroscopy of embedded 
molecules}

\vspace{2cm}

\centerline{\bf Yatendra S. Jain}

\bigskip
\centerline{\bf Department of Physics}

\smallskip
\centerline{\bf North-Eastern Hill University, Shillong - 793022, India}

\vspace{2.7cm}
\begin{abstract}
Experimental observation of superfluidity in a microscopic 
cluster, $M:(^4He)_x$, of a molecule ($M$) and $x$ number 
of $^4He$ atoms (with $x$ ranging from 1 to many) is 
qualitatively analyzed.  It concludes that: (i) each $^4He$ 
atom in the cluster has to have non-zero momentum for its 
confinement to a space of size ($<$ the size of the cluster), 
(ii) superfluidity does not require atoms with zero 
momentum ($p=0$), and (iii) while all $^4He$ atoms in the 
cluster cease to have relative motions (hence the inter-atomic 
collisions), they retain a freedom to move coherently in 
order of their locations on a closed path around the 
rotor ($M$ plus few nearest $^4He$ atoms which follow 
the molecular rotation for their relatively strong binding 
with $M$).  The analysis also identifies the basic 
arrangement of $^4He$ atoms which allows the rotor to have 
free rotation in the cluster. 

\end{abstract}



\vspace{2.7cm}

\bigskip
\noindent
\copyright $\, \,$ by author

\newpage

\bigskip
Statistical analysis of a {\it system of non-interacting bosons} (SNIB) [1] 
and Bogoliubov's field theoretical study of a system of weakly 
interacting bosons [2] provided foundations for a popular belief 
that particles in the ground state (G-state) of a {\it system of 
interacting bosons} (SIB) (such as liquid $^4He$ [3] and trapped dilute 
gases [4]) are distributed over the states of different momenta, $k=0$, 
$k_1$, $k_2$, $k_3$,  .... {\it etc.} (in wave number unit); bosons 
with zero and non-zero momenta, respectively, constitute what we call 
Bose Einstein condensate (BEC) [or zero momentum ($p$ $(=\hbar k) = 0$) 
condensate] and non-condensate.  The momentum distribution of particles 
in the G-state of a SNIB (Fig.1(A)) is compared with the above said 
distribution in a SIB (Fig.1(B)) for their better understanding. While 
the latter has been believed to exist in superfluid SIB at all 
temperatures, $T < T_c$ (the critical $T$ for the onset of superfluidity 
in a SIB), as the origin of superfluidity and related properties for the 
last seven decades, we recently discovered [5] that this distribution 
by no means represents the G-state of a SIB because it does not 
constitute a state of lowest possible energy as expected; in other 
words the laws of nature (demanding the G-state of a physical system 
to have minimum possible energy) forbids the said momentum 
distribution of bosons ($p=0$ condensate + non-condensate) 
in the G-state of a SIB.  Our study [5] also discovered the true form 
of energy/momentum distribution of particles in the G-state of a SIB, 
-accordingly, all particles in this state have to have identically 
equal energy $\varepsilon_o = h^2/8md^2$ ($\equiv$ momentum $p=h/2d$ 
depicted in Fig.1(C)) which not only represents the G-state energy (momentum) 
of a particle trapped in a cavity formed by nearest neighbors but also 
underlines the fact that not even a single particle has $p=0$;
consequently, the question of macroscopically large number of particles 
having $p=0$ does not arise.  However, the distribution (Fig.1(B)) 
seems to have strong bias in its favor, possibly, because of prolonged 
belief of people in it.  A shift from this belief, naturally, not only 
demands a theoretical foundation as discovered in [5] but also seeks 
strong experimental support for the distribution (Fig.1(C)) concluded 
in [5].  In this context, we identify several physical realities 
and experimental observations which not only support the G-state 
represented by Fig.1(C) but also refute the possibility of existence 
of $p=0$ condensate in a SIB.  We prove these points in our recent 
paper [6] (the first of a series of papers on this issue) by using 
the physical reality of the existence of an {\it electron bubble} (EB) 
in liquid helium.  Similarly, this paper (the second of the series) 
uses the experimentally observed high resolution ro-vibrational 
spectra of molecules embedded in different clusters of $^4He$ atoms 
since these spectra provide strong evidence for the resistance free 
rotations of the embedded molecule in the clusters.  

\bigskip
Ever since a systematic study [8] of the high resolution spectra 
of OCS molecule embedded in liquid $^4He$ droplets concluded 
that superfluidity is exhibited even by a drop having fewer 
(as low as 60) $^4He$ atoms, high resolution ro-vibrational 
spectra of different molecules, $M$ ($OCS$, $/N_2O$, {\it etc.}) 
embedded in $^4He$ clusters or droplets ($M:^4He_x$ where the 
number of $^4He$ atoms, $x$, changes from 1 to many), have been 
reported [9-12].  In what follows $M:^4He_x$ clusters even with 
fewer $^4He$ atoms (say 6 or so) exhibit superfluidity and the 
effective moment of inertia ($I^*$) of the molecule has 
non-trivial dependence on $x$; as expected, it first increases 
with increasing $x$ but beyond certain $x$ (depending on the 
embedded $M$) it starts decreasing with increasing $x$ and with 
further increase in $x$, it has a kind of periodic ({\it nearly})  
increase and decrease. Undoubtedly, these observations reveal 
resistance free rotational motion of the rotor ($M$ or $M$ 
attached with a few $^4He$ atoms) which underlines the fact 
that a set of $^4He$ atoms in each cluster assume the state of 
superfluid for which they do not follow the rotations of the 
rotor.  Although, numerous efforts have been made to understand 
the phenomenon, it is still not clear whether $p=0$ condensate 
of $^4He$ atoms exists in these clusters as the origin of 
the phenomenon in agreement with conventional belief and this 
motivated us to examine this issue in this paper. 

\bigskip
Since each cluster studied in these experiments has to have 
stable structure under their physical conditions, its constituents 
($M$ and $^4He$ atoms) have sufficiently strong binding 
(originating from their mutual interactions) that does not allow 
them to escape the cluster.  Further 
since their mutual interaction at short distances has infinitely 
strong repulsive character, no constituent is expected to share 
its position coordinate with others. Thus each constituent in the 
G-state of the cluster exclusively occupies 
certain space of size, $d < R$ with $R$ being the size of the 
cluster which agrees with [5, 6]; obviously, such a $^4He$ atom is 
expected to behave like a trapped particle and for this reason has 
reasonably high non-zero momentum $q = \pi/d$, certainly $ > \pi/R$.  
Evidently the question of a $^4He$ atom having zero momentum or 
the cluster having $p=0$ condensate does not arise.  It may 
be noted that, long before in 1973, Kleban [13] indicated that 
the existence of $p=0$ condensate in superfluid $^4He$ contradicts 
excluded volume condition which states that each hard 
core particle, such as $^4He$ atom, occupies certain volume 
in the fluid exclusively.  However, our analysis, reported in this 
paper and in [6], is in exact agreement with excluded volume 
condition [13].    

\bigskip
The resistance free rotation of the molecule in a cluster 
(identified as a proof of superfluid state of $^4He$ atoms) has 
no relation with $p=0$ condensate because it does not exist.  
This agrees exactly with our recent study [5] which concludes that 
the laws of nature, that demand the G-state of a SIB (microscopic 
or macroscopic) to have lowest possible energy, forbid the existence 
of $p=0$ condensate in the state.

Since this holds true for the 
G-state where superfluid density ($\rho_s$) equals the total 
density ($\rho$) of the system, possibility of the existence of 
$p=0$ condensate at non-zero $T$ where $\rho_s < \rho$ does not 
arise.   In what follows from this observation and our study [5], 
superfluidity of a cluster of $^4He$ atoms and the bulk of 
superfluid $^4He$ has a common origin.  Accordingly, it is a 
property which comes into play when all bosons assume localized 
positions with a possibility to move coherently in the order of 
their locations. 

\bigskip
When different atoms have different momenta, they have relative 
motions which render inter-atomic collisions which are expected 
to impede the rotations of the rotor in the cluster.  Hence 
the observation of free rotation of an embedded molecule in a 
cluster is a proof for the absence of relative motions of its 
constituents ($M$ and $^4He$ atoms).  Evidently, a set of such 
$^4He$ atoms would move (if they do) coherently in order of 
their locations and it is well known that atoms in superfluid 
$^4He$ really have coherence of their motion.   

\bigskip
As an important property of a fluid, its constituents move freely on 
a surface of constant potential unless they suffer mutual collisions.  
For a cluster of $^4He$ atoms, surrounding a molecule, a surface 
({\it path}) of constant potential can be a closed shell 
({\it closed orbit on a equi-potential shell}) with the molecule at 
their center.  Naturally $^4He$ atoms located on such a shell or an 
orbit would not affect a molecular rotation provided they do not 
suffer collisions which is only possible when every set of $^4He$ 
atoms move coherently in order of their location with identically 
equal momentum and two such orbits do not cross each other because 
in such a case atoms would have finite probability to collide. 

\bigskip
We believe that the physical situation offered by the cluster to its 
rotor is not much different from the situation of a molecule 
trapped at the center of a cage formed by a set of $^4He$ atoms in 
their ground state (where they have their localized positions with 
certain amount of position and momentum uncertainty) and the
interactions between the molecule and $^4He$ atoms are such that the 
molecule sees no change in its potential energy with a change in its 
angular posture with respect to stationary cage. Naturally when the 
molecule is made to rotate by its excitation it would rotate like a 
free molecule.  To be more realistic it is possible that the potential 
energy of the molecule changes with a change in its angular posture 
but with a peak value much lower than the energy of first rotational 
excitation. The shape, size and structure of the cage depend on the 
number of $^4He$ atoms which constitute a part of the rotor for their 
strong binding with the molecule due to their nearest neighbor 
positions. This number may, obviously, depend on the size of the 
embedded molecule (larger is this size, larger should be the number 
of $^4He$ atoms taking positions as its nearest neighbors). When the 
number of $^4He$ atoms available for the structure of the cage are 
very low, it could be a simple ring around the axis of rotation, 
however, with increase in the number of such $^4He$ atoms the ring 
may spread into a first shell around the rotor and with further 
increase in $x$ it may grow into several shells (second, third, ... ).  
We believe that : (i) minimum number of $^4He$ atoms available to 
form the said cage should be two, and (ii) each $^4He$ atom 
added to the cluster not only change the size, shape and structure 
of the cage, it also affects shape, size and structure of the rotor; of 
course the impact on the rotor should, obviously, diminish with growing 
size of the cage particularly after the completion of first shell. 
We note that these observations identify the basic arrangement of 
$^4He$ atoms which allows the embedded molecule to rotate like a 
free rotor.  In a recent paper [14], we used this picture to explain 
the $x$ dependence of non-trivial changes in $I^*$ of $N_2O$ and 
$HCCCN$ molecules embedded in clusters of $^4He$ atoms.    

\bigskip
Experimental observations of the resistance free rotation of a 
molecule in a cluster of $^4He$ atoms clearly refutes the 
existence of $p=0$ condensate of $^4He$ atoms.  We find that : 
(i) no $^4He$ atom in the cluster has zero momentum since it 
assumes non-zero energy and equivalent non-zero momentum for 
its confinement and this leaves no possibility for the $^4He$ 
atoms to constitute what we define as $p=0$ condensate, and (ii) 
superfluidity of $^4He$ atoms in the clusters has nothing 
to do  with $p=0$ condensate; it is a simple property which 
comes in to play because particles (molecule and $^4He$ atoms) 
in the cluster cease to have relative (collisional) motions 
(the main reason for the viscosity of a fluid) and retain the 
possibility to move in order of their locations on closed paths.  
Since our microscopic theory of a SIB identifies exactly these 
factors as the origin of superfluidity of the bulk of liquid 
$^4He$, it finds strong experimental support from the 
observation of superfluidity in the said clusters and so is 
particularly true for its conclusion about the momentum/ energy 
distribution ({\it cf.}, Fig.1(C)) of particles in the G-state 
of a SIB. With the same objective, we would study experimentally 
observed quantum evaporation of $^4He$ atoms from superfluid $^4He$ 
and the Stark effect of roton transition seen in microwave 
absorption in our forthcoming papers.

\newpage
 \bigskip
\begin{figure}
\begin{center}
\includegraphics[angle = 0, width =.6\textwidth]{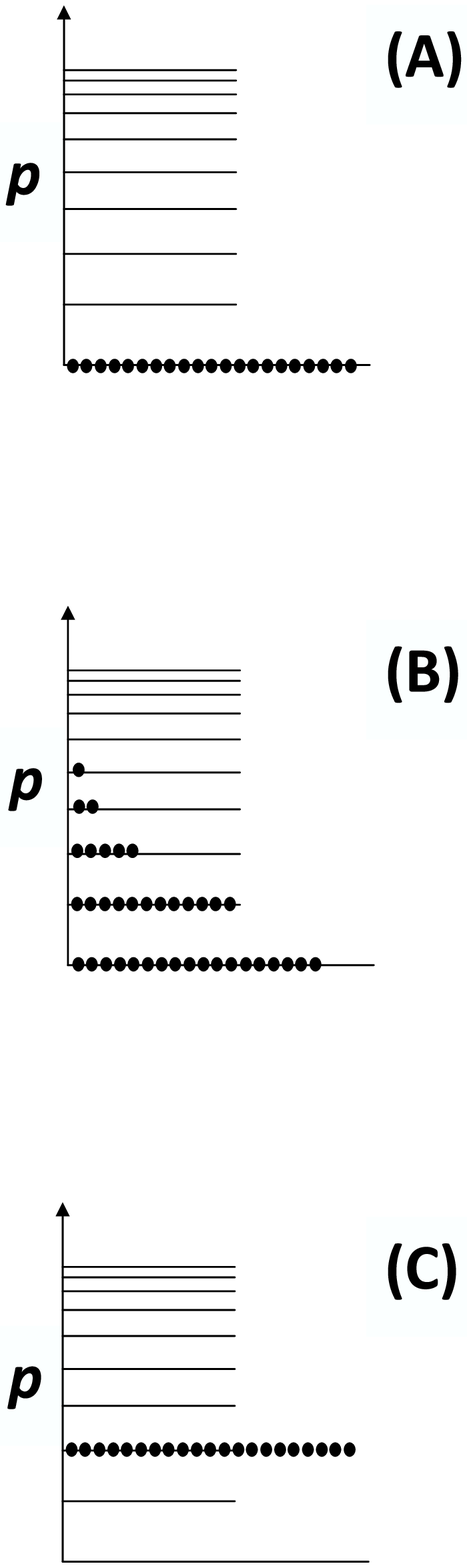}

\end{center}
\bigskip
\noindent
Fig.1 : Schematic of distribution of $N$ bosons in their 
ground state. (A) All the $N$ particles occupy $p=0$ state 
in a system of non-interacting bosons, (B) depletion of 
$p=0$ condensate ({\it i.e.} only a fraction $n_{p=0} = N_{p=0}/N$ 
of $N$ bosons occupy $p=0$ state) in weakly interacting boson 
system as predicted 
by Bogoliubov model [2], and (C) all the $N$ particles occupy 
a state of $p= p_o = \hbar q_o = h/2d$ and $\hbar K=0$ as 
concluded by this study and our recent analysis [5]. 

\end{figure}
\end{document}